\documentclass[conference]{IEEEtran}

\usepackage{epsfig}
\usepackage{graphicx}
\usepackage{amsmath}
\usepackage{amssymb}
\usepackage{amsxtra}
\usepackage{here}
\usepackage{rawfonts}
\usepackage{times}
\usepackage{url}
\usepackage{cite}

\usepackage{algorithm}

\usepackage{algorithmic}

\usepackage{adjustbox}

\usepackage{slashbox}
\usepackage{epstopdf}

\usepackage{cite}

\newtheorem{remark}{\bf Remark}
\newtheorem{theorem}{\bf Theorem}
\newtheorem{definition}{\bf Definition}

\usepackage[usenames, dvipsnames]{color}

\newlength{\aligntop}
\newlength{\alignbot}
\makeatletter

\renewenvironment{align}{%
  \vspace{\aligntop}
  \start@align\@ne\st@rredfalse\m@ne
}{%
  \math@cr \black@\totwidth@
  \egroup
  \ifingather@
    \restorealignstate@
    \egroup
    \nonumber
    \ifnum0=`{\fi\iffalse}\fi
  \else
    $$%
  \fi
  \ignorespacesafterend%
  \vspace{\alignbot}\par\noindent
}

\IEEEoverridecommandlockouts

\begin{document}
\title{\huge Data Injection Attacks on Smart Grids with Multiple Adversaries: A Game-Theoretic Perspective \vspace{-0.55cm}}
\author{\IEEEauthorblockN{Anibal Sanjab$^1$ and Walid Saad$^1$} \IEEEauthorblockA{\small
$^1$ Wireless@VT, Bradley Department of Electrical and Computer Engineering, Virginia Tech, Blacksburg, VA USA,\\
 Emails: \url{{anibals,walids}@vt.edu}\vspace{-0.8cm}
 }%
\thanks{This research was supported by the U.S. National Science Foundation under Grant CNS-1446621.}
    }
\date{}
\maketitle

\begin{abstract}
Data injection attacks have recently emerged as a significant threat on the smart power grid. By launching data injection attacks, an adversary can manipulate the real-time locational marginal prices to obtain economic benefits. Despite the surge of existing literature on data injection, most such works assume the presence of a single attacker and assume no cost for attack or defense. In contrast, in this paper, a model for data injection attacks with multiple adversaries and a single smart grid defender is introduced.  
To study the defender-attackers interaction, two game models are considered. In the first, a Stackelberg game model is used in which the defender acts as a leader that can anticipate the actions of the adversaries, that act as followers, before deciding on which measurements to protect. The existence and properties of the Stackelberg equilibrium of this game are studied.  
To find the equilibrium, a distributed learning algorithm that operates under limited system information is proposed and shown to converge to the game solution. In the second proposed game model, it is considered that the defender cannot anticipate the actions of the adversaries. To this end, we proposed a hybrid satisfaction equilibrium - Nash equilibrium game and defined its equilibrium concept. A search algorithm is also provided to find the equilibrium of the hybrid game. Numerical results using the IEEE 30-bus system are used to illustrate and analyze the strategic interactions between the attackers and defender. Our results show that by defending a very small set of measurements, the grid operator can achieve an equilibrium through which the optimal attacks have no effect on the system. Moreover, our results show how, at equilibrium, multiple attackers can play a destructive role towards each other,  
by choosing to carry out attacks that cancel each other out, leaving the system unaffected. 
In addition, we compared the obtained equilibrium strategies under the Stackelberg and the hybrid models and we characterized the amount of loss that the defender endures due to its inability to anticipate the attackers' actions.     
\end{abstract}
\vspace{-0.25cm}
\section{Introduction}
The recent integration of advanced metering infrastructures, data collection and communication nodes have rendered the smart electric grid more vulnerable to cyber-attacks. Particularly, \emph{data injection attacks} have recently emerged as an exceedingly malicious type of cyber-attacks. Using data injection, malicious adversaries can target the state estimator of a power system, by targeting a number of measurement units, in order to alter the estimate of the real-time system state~\cite{liu1stdatainjection,Poor}. 

Data injection can significantly impact the overall well-being of the power system since it can target the state estimator, an integral component of the grid which is used by the system operator to monitor, protect, control, and economically operate the system~\cite{liu1stdatainjection,Poor}. 
Using data injection attacks, malicious adversaries can achieve a variety of goals that range from compromising the security of the grid to impeding the real-time operation of the system or making financial profit through energy prices manipulation.
Data injection attacks are inherently challenging due to their stealthiness which makes the task of detecting them arduous~\cite{Poor}. 
In fact, data injection attacks can modify the estimation process while remaining unnoticed by the operator. 

Recently, data injection attacks have attracted significant attention~\cite{liu1stdatainjection,Poor,TajerPoor,LeXie,EsmalifalakGT}. The work in~\cite{liu1stdatainjection} introduces a data injection scheme that can evade detection when compromising a number of measurements. The authors in~\cite{Poor} propose an optimal data injection scheme and derive an optimized subset of measurements that can be defended to face this attack. 
The work in~\cite{TajerPoor} targets coordinated attacks and discusses efforts for detecting those attacks.  An analysis of the economic effects of data injection on energy markets is discussed in~\cite{LeXie}.
In~\cite{EsmalifalakGT}, a zero-sum game is formulated between an attacker and a defender in which the attacker modifies an estimated line flow to manipulate prices.

While interesting, this existing body of literature~\cite{Poor,liu1stdatainjection,TajerPoor,LeXie,EsmalifalakGT} has primarily focused on data injection attacks with a single attacker and assume no cost for attacking or defending the system. However, in practice, due to their potential profitability and stealthiness, data injection attacks can occur concurrently from \emph{multiple adversaries} that can target various state estimation sensors.  
Due to the networked nature of the smart grid, the manipulation of measurements in one part of the system, by an adversary, has an overall effect on the system as a whole. Hence, an attack executed by one attacker does not only impact the grid's performance, but it also affects the benefits of the other attackers. Such an interdependence can be, on the one hand, beneficial to the grid for cases in which the different simultaneous attacks mitigate the severity of one another. On the other hand, multiple attacks can lead to a more severe combined effect on the electric grid thus further impacting its overall performance. Clearly, there is a necessity for a strategic modeling framework to analyze and understand these interdependencies between attackers. Remarkably, to our best knowledge, beyond our preliminary works on data injection with two attackers in~\cite{SanjabSaad}, no work has analyzed the case of multiple adversaries.

The main contribution of this paper is to introduce novel game-theoretic approaches to analyze data injection attacks that involve a defender and \emph{multiple adversaries}. In this regard, two approaches are proposed. In the first approach, we formulate the problem as a Stackelberg game in which the defender (i.e. grid operator) acts as a leader having the ability to anticipate the actions of the adversaries, which act as followers, prior to selecting a subset of measurements to defend. The defender's goal is to reduce the effect of potential attacks on the system while optimizing a utility that captures both the benefits and costs of the chosen defense strategy. In response to the leader's strategy, the attackers play a noncooperative strategic game in which each attacker chooses its optimal attack scheme in order to maximize the trade off between the benefits, obtained from prices manipulation, and costs associated with the attack. We prove the existence of a generalized Nash equilibrium for the attacker's game and we study the existence and properties of the overall game's Stackelberg equilibrium. To solve the game, we propose a distributed learning algorithm which we prove to converge to a solution of the game using limited information that can be available to the players.  
In the second approach, it is assumed that the defender cannot anticipate the actions of the adversary. To this end, we use the framework of satisfaction equilibrium~\cite{SatEq1st} through a proposed hybrid satisfaction equilibrium - Nash equilibrium game model. In this approach, rather than anticipating the attackers' response and playing a strategy that optimizes its objective function, the defender seeks a defense strategy that meets a certain performance constraint. We introduce an equilibrium concept of this game and propose a search algorithm to find this equilibrium. 

The performance of the proposed frameworks is assessed via numerical simulations using the IEEE 30-bus test system.  
Through the numerical analysis, we simulate the strategic interactions between the attackers and defender over the test system. 
We show that by defending a minimal number of measurements, the grid operator can enforce an equilibrium in which the attacker's have no effect on the system. In addition, our results shed the light on the adversarial behavior in between the attackers. The results show that, at equilibrium, the attackers can choose attack strategies that cancel each other out resulting in no effect on the grid. In addition, we analyze the equilibrium of the hybrid game and compare the obtained solution to the Stackelberg one. In this regard, we define a ``price of information'' index which compares the utility achieved by the defender under the Stackelberg model and the hybrid model. Hence, it reflects the loss that the defender can be subject to due to lack of information about the potential reactions of the attackers to the different defense strategies available to the defender.   

The rest of this paper is organized as follows. Section~\ref{sec:model} presents the system model and problem formulation. Section~\ref{sec:GameForm} introduces the formulated Stackelberg game and associated solution. 
Section~\ref{sec:HybridSENash} introduces our proposed hybrid model and its solution concept.
Section~\ref{sec:NumRes} provides numerical results while conclusions are presented in Section \ref{sec:Conclusion}. 

\section{System Model and Problem Formulation}\label{sec:model}
\vspace{-0.1cm}
\subsection{Energy Markets}
Competitive energy markets' architectures are often based on day ahead (DA) and real time (RT) markets~\cite{PJM}. In the DA market, the system operator issues hourly-based locational marginal prices (LMPs), ${\mu^{DA}}$, for the next operating day based on the DA energy bids submitted by the participants~\cite{PJM}. 
The market clearing process is performed by the grid operator through the solution of a linearize Optimal Power Flow (DCOPF) which returns the optimal dispatch for each of the generators participating in the market and the DA LMP at each bus. The most commonly used DCOPF formulation is as follows~\cite{PJM}:\vspace{-0.2cm}

{\small\begin{align}
\label{eq:DCOPFDA}
\min_{\boldsymbol{P}}\sum\limits_{i=1}^{G}C_i(P_{i}),
\end{align}
\vspace{-0.25cm}
\begin{align}
\label{eq:pwrbalanceDA}
\textrm{s.t. } \sum\limits_{i=1}^{N}(P_i-D_i)=0,
\end{align}
\vspace{-0.25cm}
\begin{align}
\label{eq:GenLimDA}
P_i^{\textrm{min}}\leqslant P_i \leqslant P_i^{\textrm{max}},  \forall i\in\{1,\dotsb,G\},
\end{align}
\vspace{-0.5cm}
\begin{align}
\label{eq:FlowUpperDA}
\sum\limits_{i=1}^{N}(P_i-D_i)\chi_{l,i}\leqslant F_l^{\textrm{max}}, \forall l\in\{1,\dotsb,L\},
\end{align}
\vspace{-0.2cm}
\begin{align}
\label{eq:FlowLowerDA}
-\sum\limits_{i=1}^{N}(P_i-D_i)\chi_{l,i}\leqslant F_l^{\textrm{max}}, \forall l\in\{1,\dotsb,L\},
\end{align}}
\vspace{-0.3cm} \\
where $N,G$ and $L$ represent, respectively, the number of buses, generators, and transmission lines. $C_i$ corresponds to the offer of generator $i$ while $P_i$ and $D_i$ are, respectively, the power injection and load at a bus $i$. Thus, $P_i=0$ ($D_i=0$) corresponds to the case in which no generator (or load) is connected to bus $i$. The upper and lower limits on generator $i$'s output are denoted by $P_i^{\textrm{min}}$ and $P_i^{\textrm{max}}$. 
Constraints (\ref{eq:FlowUpperDA}) and (\ref{eq:FlowLowerDA}) place a limit, $F_l^{\textrm{max}}$, on the level of power that can flow over line $l$. A reference direction is assigned to the power flow over each transmission line. In this regard, a power flow opposing its assigned reference direction is represented by a negative quantity. Hence, constraints (\ref{eq:FlowUpperDA}) and  (\ref{eq:FlowLowerDA}) correspond to the thermal limit of a line in its reference and opposite directions respectively. 
$\boldsymbol{X}$ is the generation shift factor matrix which defines the sensitivity of the power flow over each line, ${\boldsymbol{F}}$, to changes in power injection, $\boldsymbol{P}$, at each bus: 
\begin{align}
\label{eq:GSF}
\boldsymbol{F}_{(L\times1)}=\boldsymbol{X}_{(L\times G)}\times \boldsymbol{P}_{(G\times1)}.
\end{align}

Therefore the sensitivity of the flow over line $l$ to a change in power injection at bus $i$ is denoted by $\chi_{l,i}$.

In the RT market, actual real-time operating conditions estimated using the state estimator, in lieu of the predictions in DA, are used through an ex-post model to compute the RT LMPs, ${\mu^{RT}}$ ~\cite{PJM}.  
An incremental DCOPF is used to compute the RT LMPs and can be formulated as follows~\cite{PJM}: 

{\small\begin{align}
\label{eq:DCOPFExpost}
\min_{\Delta \boldsymbol{P}}\sum\limits_{i=1}^{G}C_i^{RT}(\Delta P_{i}),
\end{align}
\vspace{-0.2cm}
\begin{align}
\label{eq:pwrbalanceExpost}
\textrm{s.t. } \sum\limits_{i=1}^{N}(\Delta P_i)=0,
\end{align}
\vspace{-0.2cm}
\begin{align}
\label{eq:GenLimExpost}
\Delta P_i^{\textrm{min}}\leqslant \Delta P_i \leqslant \Delta P_i^{\textrm{max}},  \forall i\in\{1,\dotsb,G\},
\end{align}
\vspace{-0.5cm}
\begin{align}
\label{eq:FlowUpperExpost}
\sum\limits_{i=1}^{N}(\Delta P_i)\chi_{l,i}\leqslant 0, \forall l\in\mathcal{C^+},
\end{align}
\vspace{-0.2cm}
\begin{align}
\label{eq:FlowLowerExpost}
-\sum\limits_{i=1}^{N}(\Delta P_i)\chi_{l,i}\leqslant 0, \forall l\in \mathcal{C^-},
\end{align}}
\vspace{-0.3cm}\\
where $C_i^{RT}$ is the RT offer of generator $i$ which is computed using its RT power output and its associated offer curve \cite{PJM}. $\mathcal{C^+}$ ($\mathcal{C^-}$) is the set of congested lines which flow is in (opposite to) their reference directions. $\Delta P_i^{\textrm{max}}$ and $\Delta P_i^{\textrm{min}}$ define a bandwidth which is employed to allow for solution tolerance. 
In practice, here, we typically \cite{LiExpostAlt} set  $\Delta P_i^{\textrm{min}}=-2$~MW and $\Delta P_i^{\textrm{max}}=+0.1$~MW. 
A proposed alternative to using this feasibility bandwidth is also available in \cite{LiExpostAlt}. 

Thus, the DA and RT LMPs at each bus, $i$, are computed using the DA and ex-post DCOPFs. The generated LMPs reflect, both, the incremental cost of energy at bus $i$ and the congestion cost associated with the contribution of this bus to the system congestion. 
A line is said to be congested if the flow of power over it reaches its maximum limit. 
The DA and RT LMPs at bus $i$ are given by:
\begin{align}
\label{eq:LMPDA}
\mu_i^{DA}=\lambda_0+\sum\limits_{l=1}^{L}(\lambda_l^{DA,-}-\lambda_l^{DA,+})\chi_{l,i},
\end{align}
\vspace{-0.2cm}
\begin{align}
\label{eq:LMPExPost}
\mu_i^{RT}=\lambda_0+\sum\limits_{l\in\mathcal{C}_l}(\lambda_l^{RT,-}-\lambda_l^{RT,+})\chi_{l,i}.
\end{align}
\vspace{-0.3cm}\\
$\mathcal{C}_l\triangleq\{\mathcal{C^+}\cup\mathcal{C^-}\}$ is the set of congested lines, in RT, obtained using the state estimator. $\mathcal{C}_l\subseteq\mathcal{L}$ where $\mathcal{L}=\{1,\dotsb,L\}$ is the set of all lines. $\lambda_0$ corresponds to the Lagrange multiplier associated with the energy balance constraints (\ref{eq:pwrbalanceDA}) 
and (\ref{eq:pwrbalanceExpost}).   
$\lambda_l^{DA,+}$ and $\lambda_l^{DA,-}$ are the Lagrange multipliers corresponding, respectively, to constraints (\ref{eq:FlowUpperDA}) and (\ref{eq:FlowLowerDA}) for line $l\in\mathcal{L}$ whereas $\lambda_l^{RT,+}$ and $\lambda_l^{RT,-}$ are the Lagrange multipliers corresponding, respectively, to constraints (\ref{eq:FlowUpperExpost}) and (\ref{eq:FlowLowerExpost}) for line  $l\in\mathcal{C}_l$.  When $l\in\mathcal{L}$ but $l\notin\mathcal{C}_l$, $\lambda_l^{RT,+}=\lambda_l^{RT,-}=0$. Moreover, when $l\in\mathcal{C}^{+}$, $\lambda_l^{RT,-}=0$;  
while when $l\in\mathcal{C}^{-}, \lambda_l^{RT,+}=0$. 

Computing the RT LMPs relies on the ex-post DCOPF formulation which depends on the output of the state estimator. Hence, data injection attacks targeting the state estimation affects the LMPs in (\ref{eq:LMPExPost}). Next, we introduce the data attack model. 
\vspace{-0.1cm}
\subsection{State Estimation and Data Injection Attacks}
\vspace{-0.1cm}
A power system state estimator uses multiple power measurements collected throughout the grid to estimate the system states~\cite{Abur}. The relation between the measurement vector, ${\boldsymbol{z}}$, and the vector of system states, $\boldsymbol{\theta}$, in a linearized state estimation model (DC SE) is expressed as follows:
\begin{align}
\label{eq:DC Model}
\boldsymbol{z}=\boldsymbol{H}\boldsymbol{\theta}+\boldsymbol{e},
\end{align}
\vspace{-0.5cm}\\
where $\boldsymbol{H}$ is the measurement Jacobian matrix and ${\boldsymbol{e}}$ is the vector of random errors assumed to follow a normal distribution, ${N(0,\boldsymbol{R})}$. Using a weighted least square (WLS) estimator the estimated system states are given by \cite{Abur}:
\vspace{-0.3cm}\\
\begin{align}
\label{eq:SE}
\boldsymbol{\hat{\theta}}=(\boldsymbol{H}^T\boldsymbol{R}^{-1}\boldsymbol{H})^{-1}\boldsymbol{H}^T\boldsymbol{R}^{-1}\boldsymbol{z}=\boldsymbol{M}\boldsymbol{z}.
\end{align}
\vspace{-0.5cm}

Using the estimated states, an estimate of the measurement vector,  $\hat{\boldsymbol{z}}$, and residuals, $\boldsymbol{r}$, can be calculated as follows \cite{Abur}:
 \begin{align}
\label{eq:zhatandresiduals}
\boldsymbol{\hat{z}}=\boldsymbol{H}\boldsymbol{\hat{\theta}}=\boldsymbol{S}\boldsymbol{z}, \ \boldsymbol{r}=\boldsymbol{z}-\boldsymbol{\hat{z}}=(\boldsymbol{I}_n-\boldsymbol{S})\boldsymbol{z}=\boldsymbol{W}\boldsymbol{z},
\end{align}
where ${I}_n$ is the identity matrix of size $(n\times n)$, and $n$ is the total number of collected measurements. 

When data injection attacks \emph{are concurrently carried out by ${M}$ attackers} in the set $\mathcal{M}=\{1,\ldots,M\}$, the collected measurements are modified through the addition of their corresponding attack vectors denoted by ${\{\boldsymbol{z}^{(1)},\boldsymbol{z}^{(2)},...,\boldsymbol{z}^{(M)}\}}$ resulting in the following altered measurements and residuals:
\vspace{+0.1cm}
\begin{align}
\label{eq:zandrwAtt}
\boldsymbol{z}^{\textrm{att}}=\boldsymbol{z}+\sum\limits_{i=1}^{M}\boldsymbol{z}^{(i)},\ 
\boldsymbol{r}^{\textrm{att}}=\boldsymbol{r}+\boldsymbol{W}\sum\limits_{m=1}^{M}\boldsymbol{z}^{(m)}.
\end{align}
\vspace{-0.3cm}

In the case in which the measurement errors $\boldsymbol{e}$ follow a normal distribution, the WLS estimator is a maximum likelihood estimator of location of the system states \cite{Abur}. 
However, the WLS estimator has a zero robustness against outliers. To overcome this drawback, outliers’ detection and identification mechanisms are used so that the final state estimate is only based on “good data”. 
The measurement residuals give an indication of the real and unknown measurement errors. 
By replacing the expression of $\boldsymbol{z}$ from (\ref{eq:DC Model}) in the expression of $\boldsymbol{r}$ in (\ref{eq:zhatandresiduals}), the residuals can be expressed in terms of the true errors as follows \cite{Abur}:
\begin{align}
\label{eq:ResidualsAndTrueErrors}
\boldsymbol{r}=\boldsymbol{W}\boldsymbol{e}
\end{align}

Thus, an analysis of the residuals allow for the detection and identification of bad data (outliers). In this respect, \emph{bad data detection} corresponds to determining whether the collected measurement set contains bad data or not. On the other hand, \emph{bad data identification} corresponds to identifying which measurements may contain bad data. One should note here that outliers can stem from data injection as well as other reasons such as meter biases or communication link failures~\cite{Abur}. 

Bad data detection is typically performed using a test known as the Chi-squares test over the sum of the squares of the residuals~\cite{liu1stdatainjection,Abur}. In fact, when the measurement errors vector $\boldsymbol{e}$ is assumed to follow a normal distribution, $||\boldsymbol{r}||_2^2=\sum_{i=1}^n r_i^2$ follows a $\chi^2$ distribution with $n-N_\theta$ degrees of freedom where $N_\theta$ is the number of states to be estimated~\cite{Abur}. Hence, for a measurement set to be considered free from bad data, the residuals must satisfy $||\boldsymbol{r}||_2\leq \tau$ where $\tau$ is a detection threshold~\cite{liu1stdatainjection,Abur}. 
In this respect, in the presence of $M$ attackers, and since $||\boldsymbol{r}^{\textrm{att}}||_2=||\boldsymbol{r}+\boldsymbol{W}\sum_{m=1}^{M}\boldsymbol{z}^{(m)}||_2$ as shown in~(\ref{eq:zandrwAtt}), each attacker $m\in\mathcal{M}$ should regulate $\boldsymbol{W}\boldsymbol{z}^{(m)}$ to keep the effect of the attacks on the residuals low to minimize the chance of being detected as outliers~\cite{liu1stdatainjection}. 

In the case where the Chi-squares test indicates the presence of bad data, various bad data identification and elimination tests can be employed such as the largest normalized residual test, or the hypothesis testing identification (HTI), among others, to identify and eliminate the outliers~\cite{Abur}.    

In our model, each attacker $m \in \mathcal{M}$ aims at manipulating RT LMPs, ${\mu^{RT}}$, to make financial benefit via virtual bidding. 
Using virtual bidding, entities that do not own any physical generation nor load can engage in the energy market settlements by submitting so-called virtual supply and demand offers. Since these energy offers are virtual, an entity offering to buy (sell) virtual power at a given bus in DA is required to sell (buy) that same amount of power at the same bus in RT. 
Using such virtual bids, the grid operator aims at promoting liquidity in the energy market while, on the other hand, virtual bidders aim to reap financial profit from possible mismatch between the DA and RT LMPs~\cite{PJM}. 
Using a data injection attack, a virtual bidder can, thus, manipulate the RT LMPs to create a lucrative mismatch with respect to their DA counterparts. On the other hand, to achieve pricing integrity, the system operator aims at protecting the system against such attacks. 
The strategic interactions between the attackers and the defender (i.e. system operator) are modeled and analyzed next.   
\vspace{-0.1cm}
\section{Attackers and Defender Strategic Interaction}\label{sec:GameForm}
Data injection attacks involve interactions between $M$ attackers, which are virtual bidders, and one defender consisting of the grid operator. 
The defender chooses a set of measurements to secure against potential attacks 
aiming at decreasing the aggregate effect of the multiple attacks on the system. Securing measurements to block data injection attacks is discussed in~\cite{Poor} and the techniques that can be implemented for securing those measurements are referred to in~\cite{Poor, EsmalifalakGT}, and~\cite{CensorProtection}. In~\cite{Poor} and~\cite{CensorProtection}, protection of measurements is performed through encryption of the associated sensors while, in~\cite{EsmalifalakGT}, protection of measurements is performed through the implementation of a set of highly secured measurement units which are assumed to provide more robustness against data attacks. 
In any case, in practice, attackers can have the ability to detect or watch which measurements are secured by the defender. 
In fact, a placement of new measurement units can be physically noticeable by the attackers while encrypting the measurement sensors' outputs can also be observed by a hacker attempting to read these outputs. 

After observing which measurements are secured, each of the $M$ attackers can choose, accordingly, to carry out a data injection attack over a subset of measurements. 
Given the networked nature of the electric grid, the actions and payoffs of the different attackers are interconnected thus motivating a game-theoretic approach \cite{GT01}.

Hence, given that the defender acts first and the attackers react to the observed defender's action, the interaction between the defender and attackers is hierarchical. Thus, we formulate a single leader, multi-follower \emph{Stackelberg game}~\cite{GT01} between the defender and the $M$ attackers to capture and analyze the strategic interaction between the two. 
In this game, the defender acts as a leader who selects a set of measurements to defend while the adversaries interact with one another using a followers noncooperative game to identify the optimal attack in response to the strategy of the defender. 
By observing or predicting the ways in which the attackers react to its defense strategy, the leader chooses its optimal defense action. 
Next, we first analyze and solve the followers game and then find the Stackelberg solution.

\subsection{Attackers' Noncooperative Game Formulation}\label{subsec:AttackersGameFormulation} 
We formulate a strategic noncooperative game to analyze the optimal decision making of the $M$ attackers in response to any arbitrary defender strategy. 
This game is formulated in its normal form as follows: $\Xi=\langle \mathcal{M}, (\mathcal{Z}^{(i)})_{i\in\mathcal{M}}, (U_{i})_{i\in\mathcal{M}}\rangle$, where $\mathcal{M}$ is the set of $M$ attackers, $\mathcal{Z}^{(i)}$ is the set of actions (attack vectors $\boldsymbol{z}^{(i)}\in \mathcal{Z}^{(i)}$) available to attacker $i \in \mathcal{M}$, and $U_i$ is the utility function of attacker $i$. 
Thus, each attacker, $m \in \mathcal{M}$, selects an attack vector, ${\boldsymbol{z}^{(m)}\in \mathcal{Z}^{(m)}}$ that maximizes its utility $U_m$. 
Let $\mathcal{K}_m$ denote the subset of measurements that $m$ can attack. Then, $\mathcal{Z}^{(m)}$ can be represented by a column vector with elements equal to 0 except for those in $\mathcal{K}_m$ which can take values within a compact range reflecting the range of magnitude of the attack.

The utility function of each attacker reflects the financial benefit obtained by virtual bidding. 
Using virtual bidding, each attacker $m$ buys and sells ${P_m}$~MW at, respectively, buses ${i_m}$ and ${j_m}$ in DA while, conversely in RT, attacker $m$ sells and buys ${P_m}$~MW at, respectively, buses ${i_m}$ and ${j_m}$. Thus, the goal of attacker $m\in{\mathcal{M}}$ is to optimize the following (Problem 1):
\vspace{-0.1cm}
\begin{flalign}
\label{eq:AttObj}
\max_{\boldsymbol{z}^{(m)}\in \mathcal{Z}^{(m)}} U_m(\boldsymbol{z}^{(m)},\boldsymbol{z}^{\textrm{$-$}(m)})\textrm{$=$}&\left[(\mu_{i_m}^{RT}\textrm{$-$}\mu_{i_m}^{DA})\textrm{$+$}\right.&\nonumber\\
 &\left.(\mu_{j_m}^{DA}\textrm{$-$}\mu_{j_m}^{RT})\right]P_m\textrm{$-$}c_m(\boldsymbol{z}^{(m)}),&
   \end{flalign}
\vspace{-0.8cm}
\begin{flalign}
\label{eq:attack_thresh}
  \textrm{s.t. }& \| \boldsymbol{W}\boldsymbol{z}^{(m)}\|_2 +\sum\limits_{l=1,l\neq{m}}^{M}\| \boldsymbol{W}\boldsymbol{z}^{(l)}\|_2 \leqslant\epsilon_m,&
\end{flalign}
where ${c_m(\boldsymbol{z}^{(m)})}$ is the cost of attack,  
and ${\boldsymbol{z}^{-(m)}}$ denotes the strategy vector of all players except $m$. The number of measurements that can be attacked concurrently by $m$ as well as the attack levels (the level of modification of a measurement) are limited by $\mathcal{Z}^{(m)}$. Since $||\boldsymbol{r}^{\textrm{att}}||_2=||\boldsymbol{r}+\boldsymbol{W}\sum_{m=1}^{M}\boldsymbol{z}^{(m)}||_2\leq||\boldsymbol{r}||_2+ ||\boldsymbol{W}\boldsymbol{z}^{(m)}||_2+\sum_{l=1,l\ne m}^{M}||\boldsymbol{W}\boldsymbol{z}^{(l)}||_2$, $m\in\mathcal{M}$ chooses $\boldsymbol{z}^{(m)}$ as in~(\ref{eq:attack_thresh}), where $\epsilon_m$ is a chosen threshold, to minimize the chance of the attack of being detected as outliers.  
\subsection{Attackers' Game Analysis}\label{subsec:AttackersGameAnalysis}
\label{subsec:GameSolution}
Due to the networked nature of the electric grid, the $m$ attackers' actions are interdependent. 
In fact, by altering a set of measurements, an attacker manipulates the whole estimation outcome and, thus, affects the actions as well as the payoffs of the other attackers. 
In the event of concurrent attacks by $M$ attackers, the resulting estimates, $\hat{\boldsymbol{z}}^{att}$, are computed as follows:
\begin{align}
\label{eq:zhatwAttackandeltazhat}
\hat{\boldsymbol{z}}^{att}=\hat{\boldsymbol{z}}+\sum\limits_{m=1}^{M}\boldsymbol{S}\boldsymbol{z}^{(m)} \Rightarrow \Delta \hat{\boldsymbol{z}}=\sum\limits_{m=1}^{M}\boldsymbol{S}\boldsymbol{z}^{(m)},
\end{align}
\vspace{-0.3cm}\\ 
where $\Delta \hat{\boldsymbol{z}}$ represents the change in the generated estimates due to the $M$ attacks. Likewise, the overall change in the measurement residuals due to the $M$ attacks can be expressed as follows:
\begin{align}
\label{eq:deltar}
\Delta\boldsymbol{r}=\boldsymbol{W}\sum\limits_{m=1}^{M}\boldsymbol{z}^{(m)}.
\end{align}\vspace{-0.35cm} \\
\indent Consequently, the various attackers in the system can impair the ability of attacker $m$ to successfully manipulate a targeted measurement $z_i$ as expressed in Remark~\ref{Rem:AttCancelation}.
\begin{remark}
\label{Rem:AttCancelation}
Depending on the targeted measurements, the collective impact of the $M$ attacks can be either constructive for the attackers by helping each one of them to achieve its goal, or destructive, attenuating the global effect of these attacks on the system.  
\end{remark}

In fact, considering the case of two attackers in which attacker $1$'s (attacker $2$'s) aim is to increase the estimated flow, $\hat{z}_i$ ($\hat{z}_j$), over a line $l_{i}$ $(l_{j})$ in order to create a false congestion. The objective of attacker $1$ (attacker $2$) is, hence, to achieve $\Delta \hat{z}_i\geqslant F^{\textrm{max}}_{l_i}-\hat{z}_i$ $(\Delta \hat{z}_j\geqslant F^{\textrm{max}}_{l_j}-\hat{z}_j)$.
Following from (\ref{eq:zhatwAttackandeltazhat}), the change introduced to $\hat{z}_i$ and $\hat{z}_j$ by the two attacks is stated as follows: 
\begin{align}
\label{eq:deltazhat2attacks}
\Delta \hat{z}_i=s_{i,i}z^{(1)}_i+s_{i,j}z^{(2)}_j, \ \Delta \hat{z}_j=s_{j,j}z^{(2)}_j+s_{j,i}z^{(1)}_i, 
\end{align}
\vspace{-0.5cm}\\
where $s_{i,j}$ denotes element $(i,j)$ of matrix $\boldsymbol{S}$. 
When the measurement errors are independent and identically distributed (i.e. $\boldsymbol{R}=\sigma^2\boldsymbol{I}_n$), $\boldsymbol{S}$ is a symmetric matrix. This property can be proven based on (\ref{eq:SE}) and (\ref{eq:zhatandresiduals}) through showing that $\boldsymbol{S}^T=\boldsymbol{S}$ when $\boldsymbol{R}=\sigma^2\boldsymbol{I}_n$.  
Since $\boldsymbol{S}$ is symmetric, ${s}_{i,j}={s}_{j,i}$. In the event where $s_{i,j}<0$, both attackers' actions attenuate the effect of one another. In fact, since ${s}_{i,j}<0$, $z^{(2)}_j$ ($z^{(1)}_i$) reduces $\Delta \hat{z}_i$ ($\Delta \hat{z}_j$) preventing it from causing any congestion over line $l_i$ ($l_j$). 
In the contrary, if $s_{ij}>0$, each of the attackers' actions would assist the other in achieving its objective. This result can be trivially generalized to the case of $M$ attackers. 

Moreover, the payoffs of the different attackers (i.e. virtual bidders) are also significantly interdependent. In fact, as shown next, an attacker can collect financial benefit or endure loses due to the strategies played by other attackers.
\begin{remark}
\label{Rem:AttUwoutAttacking}
The payoff of each attacker, $m$, is dependent on the chosen attack strategies of other attackers. Thus, based on its virtual biding nodes, $m$ can achieve a positive or negative payoff depending on attacks carried out by other attackers. 
\end{remark}

In this regard, following from (\ref{eq:AttObj}), attacker $m$'s payoff in the presence of $M$ attackers is governed by: 
\vspace{-0.3cm} \\
\begin{align}
\label{eq:attfinancialindex}
\zeta_m=(\mu_{i_{m}}^{RT}-\mu_{i_{m}}^{DA})+(\mu_{j_{m}}^{DA}-\mu_{j_{m}}^{RT}).
\end{align}
\vspace{-0.35cm}

Replacing the expressions of the DA and RT LMPs from (\ref{eq:LMPDA}) and (\ref{eq:LMPExPost}) in (\ref{eq:attfinancialindex}) yields:
{\small\begin{align}
\label{eq:attUcoupling}
\zeta_m\textrm{$=$}\sum\limits_{l=1}^L[(\chi_{l,j_m}&\textrm{$-$}\chi_{l,i_m})\textrm{$\times$} ((\lambda_l^{DA,-}-\lambda_l^{DA,+})\textrm{$+$}(\lambda_l^{RT,+}\textrm{$-$}\lambda_l^{RT,-}))].
\end{align}}
\vspace{-0.6cm}\\

As a result, following the sign of $(\chi_{l,j_m}-\chi_{l,i_m})$, determined by the choice of virtual bid nodes $i_m$ and $j_m$, an attack modifying the congestion status of a line $l$ between DA and RT can introduce a positive or negative payoff to attacker $m$.

\subsection{Attackers' Game Solution}\label{subsec:Attgamesolution}

The attackers' payoff in~(\ref{eq:AttObj}) is a function of DA and RT LMPs. These LMPs are indirectly controlled by the attack vector, $\boldsymbol{z}^{(m)}$, which can control the existence of congestion over transmission lines and hence eventually affect the LMPs in~(\ref{eq:LMPDA}) and~(\ref{eq:LMPExPost}). Thus,~(\ref{eq:AttObj}) can be rewritten using~(\ref{eq:LMPDA}) and~(\ref{eq:LMPExPost}) as:
\begin{align}\label{eq:AttObjRe}
U_m(\boldsymbol{z}^{(m)},\boldsymbol{z}^{-(m)})=\zeta_m\;P_m-c_m(\boldsymbol{z}^{(m)}),
\end{align}
where $\zeta_m$ is given by~(\ref{eq:attUcoupling}). By dropping $P_m$ for being a constant, the objective of attacker $m$ is hence to
\begin{align}\label{eq:equivalentAtt}
\max_{\boldsymbol{z}^{(m)}\in\mathcal{Z}^{(m)}}\zeta_m-c_m(\boldsymbol{z}^{(m)}).
\end{align}

We define the two sets of lines $\mathcal{L}_m^+$ and $\mathcal{L}_m^-$ such that
$\mathcal{L}_m^+=\{l\in\mathcal{L}|\chi_{l,j_m}-\chi_{l,i_m}>0\}$ and
$\mathcal{L}_m^-=\{l\in\mathcal{L}|\chi_{l,j_m}-\chi_{l,i_m}<0\}$. 
Moreover, let $\mathcal{L}^{R}$ and $\mathcal{L}^{O}$, such that $\mathcal{L}=\{\mathcal{L}^{R} \cup \mathcal{L}^{O}\}$, be the sets of lines over which the power flows in, respectively, the reference and opposite to reference directions.

Attacker $m$ seeks to congest or decongest lines in a way that maximizes~(\ref{eq:equivalentAtt}). In this regard, for a line $l\in\mathcal{L}_m^+$, i.e. $\chi_{l,j_m}-\chi_{l,i_m}>0$, attacker $m$ profits from creating a congestion over $l$ in the reference direction, causing $\lambda_l^{RT,+}$ to be positive. Thus, $m$ aims at creating a congestion over a line $l\in{\{\mathcal{L}_m^+ \cap \mathcal{L}^{R}\}}$. Similarly, for $l\in\mathcal{L}_m^-$, $m$ benefits from causing a congestion over line $l$ in the direction opposite to its reference direction, causing $\lambda_l^{RT,-}$ to be positive. Accordingly, $m$ aims a creating a congestion over a line $l\in{\{\mathcal{L}_m^- \cap \mathcal{L}^{O}\}}$. Combining these two observations, attacker $m$ aims at creating congestions over lines $l\in\{(\mathcal{L}_m^+\cap\mathcal{L}^{R})\cup(\mathcal{L}_m^-\cap\mathcal{L}^{O})\}$. In a similar manner, an attacker would also seek to remove congestion from a line $l$ in order to set its $\lambda_l^{RT,+}$ or $\lambda_l^{RT,-}$ to zero in a way that maximizes~(\ref{eq:equivalentAtt}). To this end, 
$m$ aims at removing congestions from lines $l\in\{(\mathcal{L}_m^+\cap\mathcal{L}^{O})\cup(\mathcal{L}_m^-\cap\mathcal{L}^{R})\}$.

However, due to the presence of measurement errors, an attacker cannot be completely certain that its attack will lead to the creation or removal of congestion over a given line. In fact, given the estimated states in the presence of attack, $\hat{\boldsymbol{\theta}}^\textrm{att}$, the power flow estimates, $\hat{\boldsymbol{F}}^\textrm{att}$, can be obtained using the linear matrix denoted by $\boldsymbol{H}_F$ relating the power flows to the system states: $\hat{\boldsymbol{F}}^\textrm{att}= \boldsymbol{H}_F\hat{\boldsymbol{\theta}}^\textrm{att}$.  
Using the expressions of $\hat{\boldsymbol{\theta}}$ given by~(\ref{eq:SE}), \vspace{-0.25cm}
\begin{align}
\hat{\boldsymbol{F}}^\textrm{att}=\boldsymbol{H}_F\boldsymbol{M}(\boldsymbol{z}+\sum_{i=1}^{M}\boldsymbol{z}^{(i)})=\hat{\boldsymbol{F}}+\boldsymbol{H}_F\boldsymbol{M}\sum_{i=1}^{M}\boldsymbol{z}^{(i)}.
\end{align}
\indent Replacing $\boldsymbol{z}$ by its expression given by~(\ref{eq:DC Model}) and noting that $\boldsymbol{M}\boldsymbol{H}$ reduces to the identity matrix,  
$\hat{\boldsymbol{F}}^\textrm{att}$ can be expressed as:
\begin{align}\label{eq:flowestimaterv}
\hat{\boldsymbol{F}}^\textrm{att}=\boldsymbol{H}_F\boldsymbol{\theta}+\boldsymbol{H}_F\boldsymbol{M}\boldsymbol{e}+\boldsymbol{H}_F\boldsymbol{M}\sum_{i=1}^{M}\boldsymbol{z}^{(i)}.
\end{align}
\indent $\boldsymbol{H}_F\boldsymbol{\theta}$  represents the true flow 
denoted by $\boldsymbol{F}_t$.
Given that $\boldsymbol{e}\sim N(\boldsymbol{0},\boldsymbol{R})$, $\hat{\boldsymbol{F}}^\textrm{att}$ is also a random variable that is also Gaussian distributed with the following expected value and variance: 
\begin{align}\label{eq:flowrvparameters}
\mathbb{E}[\hat{\boldsymbol{F}}^\textrm{att}]=\boldsymbol{F}_t+\boldsymbol{H}_F\boldsymbol{M}\sum_{i=1}^{M}\boldsymbol{z}^{(i)},\;V[\hat{\boldsymbol{F}}^\textrm{att}]=\boldsymbol{H}_F\boldsymbol{M}\boldsymbol{R}.
\end{align} 

Given that $\hat{\boldsymbol{F}}^\textrm{att}$ is a vector of random variables, attacker $m$ aims at altering the expected value of $\hat{\boldsymbol{F}}^\textrm{att}$ to achieve, with highest possible probability, the intended congestion creation or removal to maximize~(\ref{eq:AttObjRe}). In other words, to create (or remove) a congestion over a line $l$, attacker $m$ designs its attack so that  $\mathbb{E}[\hat{F}_l^\textrm{att}]\geqslant F_l^{\textrm{max}}+\delta_m$ (or $\mathbb{E}[\hat{F}_l^\textrm{att}]\leqslant F_l^{\textrm{max}}-\delta_m$) and aims at maximizing this $\delta_m$ to increase its chances for achieving its congestion.
Thus, attacker $m$ aims to solve the following optimization problem where $\boldsymbol{S}^F\triangleq \boldsymbol{H}_F\boldsymbol{M}$ (Problem 2):
\vspace{-0.2cm}
{\small\begin{flalign}
\label{eq:AttConvexOpti}
&\max_{\boldsymbol{z}^{(m)},\delta_{k_m},\alpha_{k_m}} \sum\limits_{k_m\in\{\mathcal{L}_m^+\cup\mathcal{L}_m^{-}\}} (\delta_{k_m}-\gamma\alpha_{k_m})-c_m(\boldsymbol{z}^{(m)})\\
&\textrm{s.t. }\>\>\| \boldsymbol{W}\boldsymbol{z}^{(m)}\|_2 +\sum\limits_{l=1,l\neq{m}}^{M}\| \boldsymbol{W}\boldsymbol{z}^{(l)}\|_2 \leqslant\epsilon_m,\ \\
&\hat{F}_{k_m}+\boldsymbol{S}^F_{k_m}\boldsymbol{z}^{(m)}+\sum\limits_{p\in\mathcal{M}\setminus{\{m\}}}\boldsymbol{S}^F_{k_m}\boldsymbol{z}^{(p)}\geqslant F_{k_m}^{\textrm{max}}+\delta_{k_m}-\alpha_{k_m}\nonumber\\ &\>\>\>\>\>\>\>\>\>\>\>\>\>\>\>\>\>\>\>\>\>\>\>\>\>\>\>\>\>\>\>\>\>\>\>\>\>\>\>\>\>\>\>\>\>\>\>\>\>\>\>\>\>\>\>\>\>\>\>\>\>\>\forall k_m\in\{\mathcal{L}_m^{+}\cap\mathcal{L}^{R}\},\\
&\hat{F}_{k_m}+\boldsymbol{S}^F_{k_m}\boldsymbol{z}^{(m)}+\sum\limits_{p\in\mathcal{M}\setminus{\{m\}}}\boldsymbol{S}^F_{k_m}\boldsymbol{z}^{(p)}\leqslant -(F_{k_m}^{\textrm{max}}+\delta_{k_m})+\alpha_{k_m}\nonumber\\ &\>\>\>\>\>\>\>\>\>\>\>\>\>\>\>\>\>\>\>\>\>\>\>\>\>\>\>\>\>\>\>\>\>\>\>\>\>\>\>\>\>\>\>\>\>\>\>\>\>\>\>\>\>\>\>\>\>\>\>\>\>\>\forall k_m\in\{\mathcal{L}_m^{-}\cap\mathcal{L}^{O}\},\\
&\hat{F}_{k_m}+\boldsymbol{S}^F_{k_m}\boldsymbol{z}^{(m)}+\sum\limits_{p\in\mathcal{M}\setminus{\{m\}}}\boldsymbol{S}^F_{k_m}\boldsymbol{z}^{(p)} \leqslant F_{k_m}^{\textrm{max}}-\delta_{k_m}+\alpha_{k_m}\nonumber\\ &\>\>\>\>\>\>\>\>\>\>\>\>\>\>\>\>\>\>\>\>\>\>\>\>\>\>\>\>\>\>\>\>\>\>\>\>\>\>\>\>\>\>\>\>\>\>\>\>\>\>\>\>\>\>\>\>\>\>\>\>\>\>\forall k_m\in\{\mathcal{L}_m^{-}\cap\mathcal{L}^{R}\},\\
&\hat{F}_{k_m}+\boldsymbol{S}^F_{k_m}\boldsymbol{z}^{(m)}+\sum\limits_{p\in\mathcal{M}\setminus{\{m\}}}\boldsymbol{S}^F_{k_m}\boldsymbol{z}^{(p)} \geqslant -(F_{k_m}^{\textrm{max}}-\delta_{k_m})-\alpha_{k_m}\nonumber\\ &\>\>\>\>\>\>\>\>\>\>\>\>\>\>\>\>\>\>\>\>\>\>\>\>\>\>\>\>\>\>\>\>\>\>\>\>\>\>\>\>\>\>\>\>\>\>\>\>\>\>\>\>\>\>\>\>\>\>\>\>\>\>\forall k_m\in\{\mathcal{L}_m^{+}\cap\mathcal{L}^{O}\},\\ \label{eq:delta}
&0<\delta_{k_m}\leqslant \beta F_{k_m}^{\textrm{max}}	 \>\>\>\> \forall \delta_{k_m}, \>\>\>
0<\alpha_{k_m}\leqslant \beta' F_{k_m}^{\textrm{max}} \>\>\> \forall \alpha_{k_m},
\end{flalign}}
where $\boldsymbol{z}^{(m)}\in\mathcal{Z}^{(m)}$. The constraints in~(\ref{eq:delta}) put some limits on the variables $\delta_{k_m}$ and $\alpha_{k_m}$ relative to the corresponding flow limit $F^{\textrm{max}}_{k_m}$ where $\beta$ and $\beta'$ correspond to the fraction of $F^{\textrm{max}}_{k_m}$ that $\delta_{k_m}$ and $\alpha_{k_m}$ can take respectively.
Thus, attacker $m$ aims at maximizing $\delta_{k_m}$ to increase the chance of creating congestions over lines $k_m\in\{(\mathcal{L}_m^+\cap\mathcal{L}^{R})\cup(\mathcal{L}_m^-\cap\mathcal{L}^{O})\}$ and removing congestions from lines $k_m\in\{(\mathcal{L}_m^+\cap\mathcal{L}^{O})\cup(\mathcal{L}_m^-\cap\mathcal{L}^{R})\}$. 

On the other hand, due to resource limitation, an attacker cannot concurrently achieve all its favorable congestions. Thus, the $\alpha_{k_m}$ variables are relaxation variables to ensure feasibility of the optimization problem. However, this relaxation is accompanied with a penalty factor, $\gamma$, present in the objective function which reflects a decrease in the objective function of the attacker for the case in which a beneficial congestion creation or removal is not performed. Hence, when such a congestion manipulation is feasible, this penalty factor ensures that the attacker has a high incentive to perform this congestion manipulation.  

Moreover, similarly to ~(\ref{eq:AttObj}), $c_m(\boldsymbol{z}^{(m)})$ is the cost associated with the attack. This cost function can be represented as a scaled norm of the attack vector, where $\kappa_m$ is the scaling factor and $m_l$ is the length of vector $\boldsymbol{z}^{(m)}$:
\begin{align}\label{eq:Attcostfunction}
c_m(\boldsymbol{z}^{(m)})=\kappa_m\sum_{i=1}^{m_l}(z^{(m)}_i)^2.
\end{align}

Following this formulation, one can see that the constraints of the optimization problems of each of the attackers are coupled. In other words, the strategy space of each attacker depends on the strategies selected by the other attackers. Games in which the constraints of the different players are coupled are known as \emph{generalized Nash equilibrium problems (GNEP)}. A widely used solution concept of these games is known as the \emph{generalized Nash equilibrium (GNE)} which is defined as follows\cite{GNEPRelaxation}:
\begin{definition}\label{Def:GNE}
In a game of $N$ players in which the control variable, i.e. strategy, of each player $i\in\{1,...,N\}$, is denoted by $\boldsymbol{x}^i\in\mathbb{R}^{n_i}$ and utility function is denoted by $U_i :\mathbb{R}^{n_1+...+n_N}\rightarrow \mathbb{R}$, a GNE is a state of the game in which each player aims at
\begin{align}
\max_{\boldsymbol{x}^i} \>\>\>U_i(\boldsymbol{x}^i,\boldsymbol{x}^{*,-i})\>\>\>
\textrm{s.t.} \>\>\> (\boldsymbol{x}^i,\boldsymbol{x}^{*,-i})\in\mathcal{X}, \label{eq:SharedX}
\end{align}
where $\boldsymbol{x}^{*,-i}$ denotes the optimal strategies of all other players except for player $i$ and $\mathcal{X}$ is the shared strategy space in between the $N$ players. In other words, as a response to optimal chosen actions of other players, a player aims at choosing the strategy, in the restricting subset dictated by the choice of the other players, that maximizes its own utility. 
\end{definition} 
  
We next prove the existence of a GNE for the attackers' game. 
\begin{theorem}\label{Theorem:AttackersGNE}
The attackers' game has at least one GNE. 
\end{theorem}
\begin{IEEEproof}
Since $\delta_{k_m}$ and $\alpha_{k_m}$  are linear functions and $-c_m(\boldsymbol{z}^{(m)})$ is a summation of strictly concave functions, as shown in~(\ref{eq:Attcostfunction}),  
each attacker's utility function given in~(\ref{eq:AttConvexOpti}) is a continuous and strictly concave function over the attackers' strategy profile. 
Moreover, $\mathcal{Z}_m$ is a convex and compact set, and as shown in~(\ref{eq:delta}), the sets in which $\delta_{k_m}$ and $\alpha_{k_m}$ lie are also compact and convex. 
Thus, since a GNEP having compact and convex action sets as well as continuous and quasi-concave utility functions has at least one GNE~\cite{Debreu1952}~\cite[Theorem 4.1]{GNEPSurvey},  
our attackers' game has at least one GNE.          
\end{IEEEproof}
The solution to GNEP problems can be obtained using a number of widely adopted solution concepts that are available in literature~\cite{GNEPSurvey,GNEPRelaxation,GNEPOptiReform} where the applicability of each technique depends on the characteristics of the utility functions and action spaces. Given the strict concavity of the utility function of each attacker's problem and the convexity of the action space, such techniques converge to a GNE for our derived formulation.
\subsection{Defender's Side Analysis}\label{subsec:DefAnalysis}
Under a given equilibrium of the followers, the leader (grid operator) selects a defense vector ${\boldsymbol{a}_0}$ that determines which measurements are to be made secure and able to block potential attacks. 
The objective of the defender is to minimize a cost function capturing the variation between the DA and RT LMPs,
on all $N$ buses in the system, as follows:
\begin{align}
\label{eq:defobj}
\min_{\boldsymbol{a}_0 \in \mathcal{A}_0} U_0(\boldsymbol{a}_0,\boldsymbol{a}_{-0})\textrm{$=$} P_L\sqrt{\frac{1}{N}\sum\limits_{i=1}^{N}(\mu_{i}^{RT}\textrm{$-$}\mu_{i}^{DA})^2}\textrm{$+$}c_0({\boldsymbol{a}_0}),
\end{align}
\vspace{-0.2cm}
\begin{align}
\label{deff_budget}
  \textrm{s.t }    \| \boldsymbol{a}_{0}\|_0\leqslant{B_0},
\end{align}
\vspace{-0.5cm}\\
where ${c_0({\boldsymbol{a}_0})}$ is the cost of defense, $P_L$ is the total system load and ${B_0}$ is the limit on the number of measurements that the operator can defend simultaneously. In~(\ref{eq:defobj}), $\mu_i^{RT}$ depends on the strategies taken by the defender, $\boldsymbol{a}_0$, and attackers, $\boldsymbol{a}_{-0}\triangleq\{\boldsymbol{z}^{(1)},\boldsymbol{z}^{(2)},...,\boldsymbol{z}^{(M)}\}$.

The Stackelberg solution concept is adequate for games with hierarchy in which the leader enforces its strategy and the followers respond, rationally (i.e. optimally), to the leader's strategy~\cite{GT01}. We denote the optimal response of the attackers to action $\boldsymbol{a}_0$ played by the defender by $\mathcal{R}^{\textrm{att}}(\boldsymbol{a}_0)\triangleq\{\boldsymbol{z}^{{(1)}^*}(\boldsymbol{a}_0), \boldsymbol{z}^{{(2)}^*}(\boldsymbol{a}_0),\dotsb, \boldsymbol{z}^{{(M)}^*}(\boldsymbol{a}_0)\}$. This optimal strategy denotes the equilibrium strategy profile of the attackers as a response to the defender's strategy. In this regard, $\boldsymbol{a}_0^*\in\mathcal{A}_0$ is a Stackelberg equilibrium if it minimizes the leader's (i.e. defender's) utility function $U_0$. In other words,

\vspace{-0.3cm}
\begin{align}\label{eq:StackelbergDef}
U_0(\boldsymbol{a}^*_0, \mathcal{R}^{\textrm{att}}(\boldsymbol{a}^*_0)) \leqslant U_0(\boldsymbol{a}_0, \mathcal{R}^{\textrm{att}}(\boldsymbol{a}_0)) \,\,\forall \boldsymbol{a}_0\in\mathcal{A}_0.
\end{align}  

A Stackelberg equilibrium is guaranteed to exist and be unique if the optimal response of the followers is unique in response to every action of the leader. 
However, Theorem~\ref{Theorem:AttackersGNE} proves the existence of at least one GNE for the attackers' game. Hence, the followers can have multiple optimal responses to a leaders strategy. In this case, the leader can rank the GNEs corresponding to each strategy based on their impact on its utility and retain the one that leads to the worst utility (i.e. maximal utility given that the defender is a utility minimzer). The leader then selects the policy that minimizes this maximal utility. This is known as a hierarchical equilibrium (HE)~\cite{GT01}. In other words, $\boldsymbol{a}_0\in\mathcal{A}_0$ is a hierarchical equilibrium strategy for the defender if:
\begin{align}\label{eq:minmax}
{\small
\max_{\boldsymbol{a}_{-0}\in\mathcal{R}^{\textrm{att}}(\boldsymbol{a}^*_0)} U_0(\boldsymbol{a}^*_0,\boldsymbol{a}_{-0})=\min_{\boldsymbol{a}_0\in\mathcal{A}_0} \max_{\boldsymbol{a}_{-0}\in\mathcal{R}^{\textrm{att}}(\boldsymbol{a}^0)} U_0(\boldsymbol{a}_0,\boldsymbol{a}_{-0}).}
\end{align}     

\vspace{-0.1cm}
\subsection{Distributed Learning Algorithm}\label{sec:Learning}

Here, we provide a methodology for finding a hierarchical equilibrium of the defender-attackers game as defined by~(\ref{eq:minmax}). 

We first consider the attackers subgame. To find an equilibrium that can be reached by the attackers, we propose a distributed learning algorithm, based on the framework of learning automata that was first analyzed in ~\cite{AutomatonBook}. 
The main drivers behind this algorithm are as follows. First, this algorithm is fully distributed in the sense that each attacker is only required to know its own action space, and not the shared one, and the observation of its own payoff after choosing an action. In this regard, knowledge of the action spaces of other attackers or even their existence is not required. 
Second, since the attackers' game might admit multiple GNEs, the use of this algorithm, emulating practical smart grids security settings, enables the characterization of the GNE(s) that can be actually reached in practice.

The proposed learning algorithm is shown in Algorithm~\ref{alg:alg1}. In this algorithm, each attacker $m$ first initializes a strategy vector $\boldsymbol{q}^{(m)}$ containing a probability distribution over its attack space\footnote{This algorithm requires decritization of the action space of the attackers; our discretization approach is provided in Section~\ref{sec:NumRes}.}. For instance, $q^{(m)}_{\boldsymbol{z}^{(m)}}(t)$ corresponds to the probability that attacker $m$ chooses attack $\boldsymbol{z}^{(m)}$ at time instant $t$. Then, at time instant $t$, each attacker chooses an attack randomly and independently from its attack space following the probability distribution available through its strategy vector. The collection of the attackers' actions at time instant $t$ results in a payoff for each attacker denoted by $r_m(t)$. $r_m(t)$ is a positive normalized value which corresponds to a mapping from $[U_m^{\textrm{min}},U_m^{\textrm{max}}]\rightarrow[0,1]$ where $U_m^{\textrm{min}}$ and $U_m^{\textrm{max}}$ are the minimum and maximum achievable utilities by $m$. 
Based on the payoff that it receives at time instant $i$, each attacker, $m\in\mathcal{M}$, updates its strategy vector as follows:
\begin{align}\label{eq:automaton}
\boldsymbol{q}^{(m)}(t+1)=\boldsymbol{q}^{(m)}(t)+b\,r_{m}(t)(\boldsymbol{e}^{(m)}(t)-\boldsymbol{q}^{(m)}(t)) ,
\end{align}   
where $b$ is an arbitrarily small positive constant and $\boldsymbol{e}^{(m)}(t)$ is a column vector of length equal to the size of the action set of attacker $m$. $\boldsymbol{e}^{(m)}(t)$ is a vector whose elements are equal to 0 except for the element corresponding to the action that was selected at time instant $t$. The element corresponding to the selected action will have a value of 1. Thus, given that the $j^{th}$ attack was selected by attacker $m$ at time instant $t$; then, $e^{(m)}_j(t)=1$ and $e^{(m)}_k(t)=0$ for $k\neq j$. This updating scheme is known as a \emph{linear reward-inaction ($L_{R-I}$)} scheme~\cite{AutomatonBook}. Hence, with every iteration, the strategy vector of each attacker is updated and the algorithm repeats until each of the attackers' strategy vectors has all elements equal to 0 except for one element which is equal to 1. Such a strategy vector shows which of the strategies is to be chosen by each attacker. The collection of these attacks (having a probability of 1 each) corresponds to the game's equilibrium.    
\begin{algorithm}[!t]
\caption{Distributed Learning Automata}
\label{alg:alg1}
\begin{algorithmic}[1]
\REQUIRE Number of attackers $M$
\\\setlength\parindent{15pt} Action space of each attacker $\mathcal{Z}^{(m)}$
\ENSURE Strategy vector of each player $\boldsymbol{q}^{(m)}$
\STATE Initialize $\boldsymbol{q}^{(m)}(0)$
\WHILE {Not Converged}
\STATE Randomly select $\boldsymbol{z}^{(m)}(t)$ based on $\boldsymbol{q}^{(m)}(t)$
\STATE Collect payoff $r_m(t)$
\STATE Update strategy vector
\\ $\boldsymbol{q}^{(m)}(t+1) = \boldsymbol{q}^{(m)}(t) + b\,r_m(t)\left(\boldsymbol{e}^{(m)}(t)-\boldsymbol{q}^{(m)}(t)\right)$
\STATE Check Convergence
\IF {Converged}
\STATE{ Break}
\ENDIF
\ENDWHILE
\RETURN Strategy vector $\boldsymbol{q}^{(m)}$ 
\end{algorithmic}
\end{algorithm}
This algorithm has been discussed in\cite{AutomatonBook, AutomatonPaper} where it has been proven that, for an arbitrarily small $b$, this algorithm asymptotically converges to a pure strategy Nash equilibrium (PSNE) when the game admits a PSNE.
\begin{definition}\label{Def:PSNE}
Following the notations of Definition~\ref{Def:GNE}
a PSNE is a state of the game in which each player aims at
\begin{align}
\max_{\boldsymbol{x}^i} \>\>\>U_i(\boldsymbol{x}^i,\boldsymbol{x}^{*,-i})\>\>\>
\textrm{s.t.} \>\>\> \boldsymbol{x}^i\in\mathcal{X}_i \label{eq:NonSharedX},
\end{align}
where, on the contrary with GNEP, $\mathcal{X}_i$ is player $i$'s own strategy space which is independent of other players. 
\end{definition}

Thus, the main difference between a GNE and a PSNE is that the GNE is an optimal action profile in which each action does not violate coupled constraints with other players. Thus, given that Algorithm~\ref{alg:alg1} is guaranteed to converge to a PSNE, proving that this PSNE will never violate the coupled constraints is enough to prove the convergence to a GNE.
\begin{theorem}\label{Theorem:AutomataConvergenceGNE}
When applied to Problem 1, Algorithm~\ref{alg:alg1} is guaranteed to asymptotically converge to a GNE 
when the step size $b$ is chosen to be arbitrarily small. 
\end{theorem}
\begin{IEEEproof}
For a strategy $\boldsymbol{z}^{(m)^*}$ to be a best response strategy (BR) for attacker $m$, it needs to satisfy the property $U_m(\boldsymbol{z}^{(m)^*},\boldsymbol{z}^{-(m)})\geq U_m(\boldsymbol{z}^{(m)},\boldsymbol{z}^{-(m)}) \, \forall \boldsymbol{z}^{(m)}\in\mathcal{Z}^{(m)}$. A PSNE is hence a state of the game in which all players play BR strategies with respect to one another. Thus, a strategy that is not a BR strategy cannot be a PSNE strategy. 

However, a strategy $\boldsymbol{z}^{(m)}$ that violates the residual threshold constraints in~(\ref{eq:attack_thresh}) cannot be a BR strategy. In fact, consider the case in which attacker $m$ attacks only one measurement $z_i$ and its attack is denoted by $z^{(m)}_i$. If this attack violates the residual threshold of $z_i$, $z_i$ will be identified as outlier and discarded from the measurement set. 
Thus, from~(\ref{eq:equivalentAtt}), this results in
\begin{flalign}
U_m(z^{(m)}_i,\boldsymbol{z}^{-(m)})&=\zeta_mP_m-c_m(z^{(m)})\nonumber\\
&<\zeta_mP_m=U_m(0,\boldsymbol{z}^{-(m)}).\nonumber
\end{flalign}

Thus, $z^{(m)}$ is not a BR since not launching an attack at all returns a higher $U_m$. 
Hence, all actions that violate the coupled constraints are dominated by the alternative of not carrying out an attack at all and hence cannot correspond to BR strategies. 
As a result, a PSNE is guaranteed not to violate the coupled constraints in~(\ref{eq:attack_thresh}) and hence this PSNE is a GNE of the game. 
Since all PSNEs are GNEs and that algorithm~\ref{alg:alg1} asymptotically converges to a PSNE for a small $b$ it, as a result, converges to a GNE of our game.   
\end{IEEEproof}

To be able to choose a hierarchical equilibrium strategy, a defender needs to anticipate the worst case GNE of the attacker to each defense strategy. This anticipation can be done through: i) repeating the learning algorithm of the attackers, Algorithm 1, starting from different initial conditions to find all possible GNEs from which worst case GNEs can be extracted, ii) using one of the various algorithms tailored to find solutions of GNE problems~\cite{GNEPSurvey,GNEPRelaxation,GNEPOptiReform}, iii) analytical derivation based on its full knowledge of the cyber-physical system model, energy market model and available past data.

When being able to anticipate all worst case GNEs of the attackers, the defender chooses the strategy that results in the best worst case GNE. This strategy and its corresponding GNE corresponds to the HE of the game.\vspace{-0.2cm}
\section{Game model under limited information}\label{sec:HybridSENash}
Thus far, we assumed that the defender can anticipate all worst case GNEs of the attackers. However, in some instances, the defender does not have enough knowledge to anticipate the reaction of the attackers. Thus, it cannot seek a strategy that minimizes its utility when the reaction of the followers to any of its actions is unknown. As a result, we employ the framework of \emph{satisfaction equilibrium} (SE)~\cite{SatEq1st,SatEqDebbah}. 
Under the satisfaction framework, rather than minimizing its objective function~(\ref{eq:defobj}), given a number of measurements that can be defended concurrently, $b_0$, the defender aims at keeping the overall changes in the LMPs at all buses under a desired threshold $\Gamma_0$: \vspace{-0.05cm} 
\begin{align}\label{eq:DefSatFc}
r_0(\boldsymbol{a}_0,\boldsymbol{a}_{-0})=\sum\limits_{i=1}^{N}(\mu_{i}^{RT}-\mu_{i}^{DA})^2 \leq \Gamma_0.
\end{align} 
\indent Given our hierarchical model, we present a hybrid SE-Nash model in which the defender aims at choosing an action that satisfies its performance requirement given potential reaction of the attackers while the attackers observe the action of the defender and play a noncooperative game in which each attacker aims at maximizing its utility. Extending the SE logic to the attackers game, an attacker is satisfied by playing one of its BR strategies facing the actions chosen by other attackers and the defender.   
We denote an equilibrium of this hybrid model as a hybrid hierarchical equilibrium (HHE).
\begin{definition}\label{Def:HybridSat-Nash}
A strategy profile $(\boldsymbol{a}_0^*,\boldsymbol{z}^{(1)^*},...,\boldsymbol{z}^{(M)^*})$ is an HHE if $r_0(\boldsymbol{a}_0^*,\boldsymbol{a}_{-0}^*)\leq\Gamma_0$ and $U_m(\boldsymbol{z}^{(m)^*},\boldsymbol{z}^{-(m)^*})\geqslant U_m(\boldsymbol{z}^{(m)},\boldsymbol{z}^{-(m)^*})\, \forall \boldsymbol{z}^{(m)}\in\mathcal{Z}^{(m)}$ and $m\in\{1,...,M\}$. 
\end{definition}

With a proper choice of $\Gamma_0$ this game is guaranteed to have at least one HHE. In fact, if none of the actions available to the defender leads to meeting its satisfaction level then either the satisfaction threshold needs to be increased or more resources should be employed so that a larger number of measurements can be concurrently secured. When the leader chooses an action that satisfies~(\ref{eq:DefSatFc}) it has no incentive to deviate from it. The attackers will respond to this strategy by playing a GNE. Hence, the attackers would also have no incentive to deviate from this GNE. As a result, the satisfaction strategy of the defender and its GNE response by the attackers correspond to an HHE of this hybrid game.
  
Given the lack of knowledge about the adversaries, the defender has to learn the action(s) that insure the satisfaction of its constraint through trial and observation. To this end, to find a strategy that satisfies its performance constraint, the defender can adopt the following search algorithm:
\begin{itemize}
\item[i)] For a maximum number of iterations, $N_0$, the defender starts by choosing an action from its strategy space $\mathcal{A}_0$ following a uniform probability distribution $\boldsymbol{f}_0$ over this action space. The followers observe this action and react by playing a noncooperative game whose GNE is obtainable via Algorithm 1. 
\item[ii)] The leader observes if the action it had taken led to the satisfaction of its performance constraint. If this is the case, the strategy is hence deemed satisfactory and the leader has no incentive to deviate from it. The followers response to the leader's satisfaction action is a GNE and hence the followers have no incentive to deviate from their response as well. Thus, this results in an equilibrium. 
\item[iii)] If the action that the defender had chosen did not lead to the satisfaction of its constraint, another action is randomly chosen from its strategy space and the process repeats.
\end{itemize}

This algorithm will eventually find a HHE since, for a large number of iterations and given that at least one action exists in its action space that satisfies the defender's threshold, this action would eventually be randomly chosen with a probability that is extremely close to 1. Assume the number of vulnerable measurements to be equal to $V$ and that the defender secures $b_0<V$ measurements concurrently. Its action space has then a cardinality $|\mathcal{A}_0|=V!/(b_0!(V-b_0)!)$. Assume that $n_0$ of the alternatives achieve $r_0\leq\Gamma_0$. Choosing uniformly between the alternatives, the probability of choosing an action that satisfies the defender is equal to $p_0=n_0/|\mathcal{A}_0|$. Thus, the probability of not finding a satisfaction action in $N_0$ iterations, i.e. trials, is given by $(1-p_0)^{N_0}$; and hence, the probability of finding a satisfaction action in $N_0$ iterations 
is given by $p_0^*=1-(1-p_0)^{N_0}$. Moreover, the expected number of iterations needed to find a satisfaction equilibrium strategy is equal to $\mu_0=1/p_0$ while the variance of that number is equal to $v_0=(1-p_0)/p_0^2$.  
\indent Hence, significantly increasing $\Gamma_0$ will typically increase $n_0$ leading to an increase in $p_0$ and $p_0^*$ and a decrease in $\mu_0$ and $v_0$. As a result, one can see the conflicting effect between the quality of the found solution, reflected by how low the satisfaction threshold $\Gamma_0$ is, and the speed of finding a solution. Reducing the required satisfaction quality leads to finding a solution faster while a higher satisfaction quality requirement (lower $\Gamma_0$) leads to a slower identification of a solution.\vspace{-0.2cm}
\section{Numerical Results and Analysis}\label{sec:NumRes}
For performance evaluation, we consider three data injection attackers and one defender interacting over the IEEE 30-bus test system which represents a segment of the American Electric Power System \cite{30BusAlsac,30BusFerrero}.    

In our numerical setting, each attacker is assumed to have a subset of measurements comprising three measurements that it can attack. In particular, attacker 1 can attack line flow measurements over lines 3, 4 and 7, attacker 2 can attack line flow measurements over lines 14, 15 and 16, and attacker 3 can attack line flow measurements over lines 5, 9 and 11. The attack level on any of the measurements is assumed to have one of the following power levels (in MW): $\{-3.5, 2, 0, 2, 3.5\}$. The amount of virtual power that each attacker sells or buys is assumed to be equal to 100 MW and its attack cost is as shown in (\ref{eq:Attcostfunction}) where $\kappa_m=0.25$.
\begin{table}[t!]
\caption{Attackers' Virtual Bidding Configuration \label{table:AttInfo}} 
\begin{center}
\begin{tabular}[b]{|l|c|c|c|}
\hline
{\footnotesize{Attacker}} & {VB Bus 1} & {VB Bus2} & {Target Line} \\
\hline
{Attacker 1} & {Bus 3} & {Bus 4} & {Line 4}\\
\hline
{Attacker 2} & {Bus 4} & {Bus 12} & {Line 15}\\
\hline
{Attacker 3} & {Bus 6} & {Bus 7} & {Line 9}\\
\hline
\end{tabular}
\end{center}
\vspace{-0.6cm}
\end{table}
The DA and RT virtual bidding (VB) information of the different attackers are shown in Table~\ref{table:AttInfo}. In this table, VB Bus 1 corresponds to the bus at which an attacker sells energy in DA (respectively buys in RT) and VB Bus 2 corresponds to the bus at which this attacker buys energy in DA (respectively sells in RT). The target line column corresponds to the line connecting the two VB buses which the attacker aims to congest.      
In our simulations, we assume that the system experiences no congestion in DA and that each attacker primarily aims at creating a fake estimated congestion over its target line so as to reap financial benefit\footnote{Since our main focus is on the attackers' and defender's strategies, it is assumed that all market participants abide by their DA schedules and, except for the attacks and defense, no change in system conditions occurs between DA and RT. Thus, in case of no attacks, the DA and RT LMPs match.}.

On the other hand, the defender decides on a subset of measurements to secure out of all the measurements in the system. In our simulations, we assume that a measurement device is placed on every bus and every line in the system so that every power injection and every line flow is measured. 

\begin{figure}[t!]
  \begin{center}
   \vspace{-0.35cm}
    \includegraphics[width=8cm]{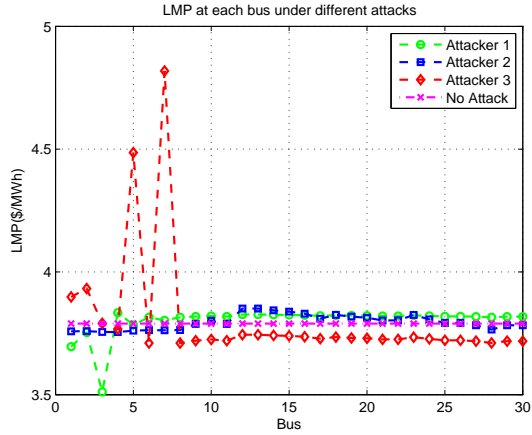}
   \vspace{-0.35cm}
    \caption{\label{fig:LMPswDifferentAttacks} System LMPs under a unique adversary's optimal attack with no defense}
  \end{center}\vspace{-0.35cm}
\end{figure}

\begin{figure}[t!]
  \begin{center}
   \vspace{-0.35cm}
    \includegraphics[width=7.5cm]{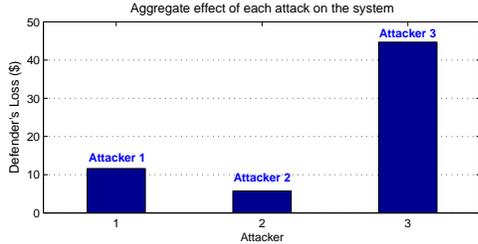}
    \vspace{-0.35cm}
    \caption{\label{fig:DefenderCostDiffAttacks} Global effect of each attack on the system.}
  \end{center}
\end{figure}

In Fig.~\ref{fig:LMPswDifferentAttacks} we show the effect of each attacker's optimal attack, when no defense or other attackers are present in the system, on the RT LMPs.   
As can be seen from Fig.~\ref{fig:LMPswDifferentAttacks}, assuming that only one attacker attacks at a time, the action of attacker 3 yields the most detrimental effect on the system. This can be also seen from Fig.~\ref{fig:DefenderCostDiffAttacks} in which the impact of each of the attacks on the system is shown. The global effect of any attack on the system is captured by the defender's utility function given in~(\ref{eq:defobj}). Fig.~\ref{fig:DefenderCostDiffAttacks} shows indeed that the attack of 3 has the highest global effect on the system followed by that of attacker 1 and of attacker 2 respectively. In fact, the defender's loss under the attack of attacker 3 is equal to \$44.69 as compared to \$11.61 under that of attacker 1 and \$5.74 under that of attacker 2.  

\begin{figure}[t!]
  \begin{center}
   \vspace{-0.35cm}
    \includegraphics[width=8cm]{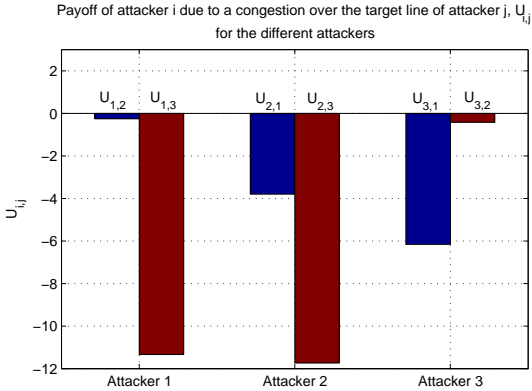}
    \vspace{-0.35cm}
    \caption{\label{fig:Uij} Loss to attacker $i$ due to congestion over the target line of attacker $j$.}
  \end{center}\vspace{-0.8cm}
\end{figure}

Fig.~\ref{fig:Uij} shows the effect that a congestion over the target line of attacker $j$ has on the payoff of attacker $i$ denoted as $U_{i,j}$. Accordingly, Fig.~\ref{fig:Uij} shows how the attack of an attacker affects the payoffs of the others. To this end, the attackers appear to be in a perfectly conflicting situation since fulfilling the purpose of an attacker $i$ results in a negative payoff to all other attackers.   

Next, we consider the strategic interactions between the three attackers and the defender based on our Stackelberg model. In this regard, we find the HE of the game, and the underlying GNE of the attackers, when the defender defends an increasing number of measurements as shown in Table~\ref{table:Game Solution}. We, namely, treat the cases in which the number of measurements that can be defended concurrently, $B_0$, is $0$, $1$ and $2$. The attackers' optimal strategies are represented in a vector containing their respective optimal attack levels (in MW) such that for attacker 1 the attacked levels correspond to additive power flows over $(\textrm{line}\,3, \textrm{line}\,4, \textrm{line}\,7)$, for attacker 2 to additive power flows over $(\textrm{line}\,14, \textrm{line}\,15, \textrm{line}\,16)$ and for attacker 3 to additive power flows over $(\textrm{line}\,5, \textrm{line}\,9, \textrm{line}\,11)$.
 
Given that a congestion occurring over line $9$ has the largest impact on the system, one can intuitively expect the defender to secure the measurement over that line when only one measurement can be secured (i.e. $B_0=1$). Indeed, from Table~\ref{table:Game Solution}, we can see that the HE corresponds to the defender defending line $9$ and the attackers carrying out their optimal equilibrium response. For the case in which the defender can defend up to two measurements concurrently, i.e. $B_0=2$, one expects the defender to secure the measurements of the two lines, lines 4 and 9, which congestion has the largest impact on the system (we refer to this defense, in this context, as the critical defense). However, the HE of the game corresponds to the defender defending lines 4 and 5 instead. In fact, by defending those two lines \emph{the attackers' optimal response yields no effect on the system hence leaving the RT LMPs unaffected}. This is a representation of the analysis provided through Remark~\ref{Rem:AttCancelation} in which multiple attackers' attacks can cancel each other out.    
\begin{table}[t!]
\caption{Stackelberg Game Solution for $B_0=\{0,1,2\}$ \label{table:Game Solution}} \vspace{-0.55cm}
\begin{center}
\begin{tabular}[b]{|l|c|c|c|c|}
\hline
{\footnotesize{$B_0$}} & {Secured Measurements} & {Attacker 1} & {Attacker 2} & {Attacker 3} \\
\hline
{0} & {-} & {(2,3.5,3.5)} & {(0,0,0)} & {(3.5,3.5,0)}\\
\hline
{1} & {line 9} & {(0,3.5,2)} & {(0,0,0)} & {(0,0,0)}\\
\hline
{2} & {lines 4 and 5} & {(0,0,0)} & {(0,2,0)} & {(0,3.5,-3.5)}\\
\hline
\end{tabular}
\end{center}
\vspace{-0.3cm}
\end{table}
\begin{figure}[t!]
  \begin{center}
   \vspace{-0.35cm}
    \includegraphics[width=7.8cm]{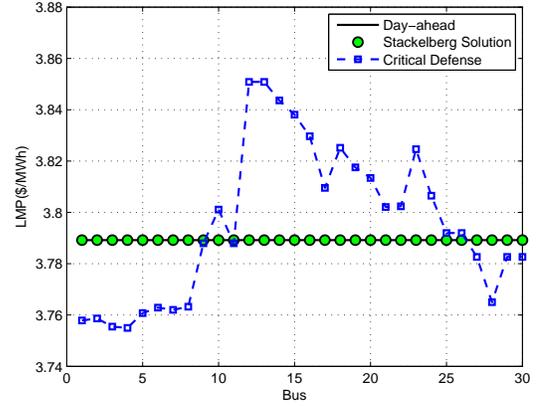}
    \vspace{-0.4cm}
    \caption{\label{fig:Def2lines} Comparison between HE and critical line defense strategies}
  \end{center}\vspace{-1cm}
\end{figure}

\begin{figure}[t!]
  \begin{center}
   \vspace{-0.35cm}
    \includegraphics[width=8cm]{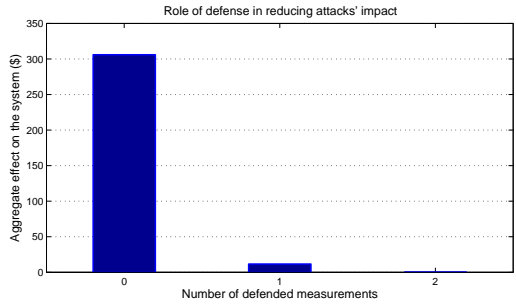}
    \vspace{-0.35cm}
    \caption{\label{fig:DefenseRole} Effect of system defense}
  \end{center}\vspace{-0.9cm}
\end{figure}

Fig.~\ref{fig:Def2lines} provides a comparison between the LMP manipulation outcome under the HE strategy as compared to the critical defense strategy. It can be clearly seen that the HE strategy (i.e. Stackelberg solution) completely prevents the manipulation of the RT LMPs and, hence, is a significantly better strategy than critical defense. In fact, the attackers' optimal response to the critical defense strategy resulted in a successful manipulation of the RT LMPs leading to a $3\%$ root mean square deviation (RMSD) from the DA LMPs where 
\begin{align}\label{eq:RMSD}
{\small
\textrm{RMSD}=\sqrt{\frac{1}{N}\sum\limits_{i=1}^{N}(\mu_{i}^{RT}-\mu_{i}^{DA})^2}.}
\end{align}
\\\vspace{-0.05cm}\indent 
Fig.~\ref{fig:DefenseRole} shows the defender's HE utility for different numbers of concurrently defended measurements. This figure shows that the attackers have a very large impact on the system when no defensive actions are taken. In fact, the aggregate effect of the attacks on the system LMPs, at equilibrium, with no defensive actions is \$306.  However, when $B_0=1$ (defender is able to secure one measurement), the HE of the game shows that the global effect of the multiple data injections attacks on the system drops significantly to \$11.6. Moreover, when $B_0=2$ (defender is able to secure two measurements concurrently), the HE of the game shows that the defender's equilibrium strategy completely protects the system against attacks and achieves a zero overall effect of the attacks on the system.   
 
We next consider our proposed SE-Nash framework and we define the \emph{price of information} (PI) to be an index reflecting the loss that the defender endures due to its lack of information about the possible reaction of the attackers to its defense strategies. The PI is defined as follows:
\begin{align}
\textrm{PI}=U_0^\textrm{HHE}-U_0^\textrm{HE}.
\end{align}    
The PI hence reflects the difference between the utility achieved under the SE-Nash framework and the one achieved under the Stackleberg model (corresponding to minimum possible utility).

We consider first that the defender can only defend one measurement at a time. Given that there are 9 vulnerable measurement units in the system, the defender has 9 options to choose from. Considering that the defender would be satisfied by having $\textrm{RMSD}\leq10\%$, we run the search algorithm described in Section~\ref{sec:HybridSENash} and the HHE we obtained is similar to the one we obtained using the Stackelberg model for $B_0=1$. Thus, in this case, $\textrm{PI}=0$. Following this HHE, the defender defends the line measurement over line $9$ and the attackers' GNE corresponds to that shown in Table~\ref{table:Game Solution} for $B_0=1$. This strategy generates an $RMSD=6.1\%<10\%$ hence meeting the performance requirement. 

Next, we consider the case in which the defender secures 2 measurements concurrently. Thus, the defender has 36 options to choose from. We consider two different performance requirements. In the first, the defender seeks to have $\textrm{RMSD}\leq5\%$. Running the search algorithm yields an HHE dictating the defense of lines $4$ and $9$ which corresponds to the critical defense defined previously. This HHE results in $\textrm{RMSD}=3\%<5\%$ and a $\textrm{PI}=\$5.74$. The second considered performance requirement seeks to have $\textrm{RMSD}\leq10\%$. In this regard, our search algorithm led to an HHE under which lines $5$ and $9$ should be defended. This HHE results in $\textrm{RMSD}=6.1\%<10\%$ and a $\textrm{PI}=\$11.61$.
\section{Conclusion}\label{sec:Conclusion}
\vspace{-0.08cm}
In this paper, we have studied the problem of data injection attacks on the smart grid in the presence of multiple adversaries. The strategic interactions between the defender and the attackers have been modeled using a Stackelberg game and a hybrid satisfaction equilibrium - Nash equilibrium game. In these games, the grid operator acts as the leader and the attackers act as followers which play a noncooperative strategic game in response to each defender's strategy. The costs of attack and defense have been integrated in the utility functions of the players. We have proven the existence of a generalized Nash equilibrium of the attackers' game, studied the existence and properties of the equilibria of the Stackelberg and the hybrid games and proposed learning algorithms, and proved their convergence, to compute the games' solutions. Numerical results have shown the critically important role of the defender in protecting the grid and the potential conflicting interaction between the multiple adversaries. Our results also highlight potential loss that the defender can incur due to a lack of information about the actions of the attackers. \vspace{-0.2cm}
\def\baselinestretch{0.85}
\bibliographystyle{IEEEtran}
\bibliography{reference}

\end{document}